# CLUES ON THE IMPORTANCE OF COMETS IN THE ORIGIN AND EVOLUTION OF THE ATMOSPHERES OF TITAN AND EARTH

**Josep M[a] Trigo-Rodriguez[1,2]**

[1] Institute of Space Sciences –CSIC, Campus UAB, Facultat de Ciències,
Torre C5-parell-2[a], 08193 Bellaterra, Barcelona, Spain. E-mail: trigo@ieec.uab.es
[2] Institut d'Estudis Espacials de Catalunya (IEEC), Edif.. Nexus,
c/Gran Capità, 2-4, 08034 Barcelona, Spain

**and**

**F. Javier Martín-Torres**
Centro de Astrobiología (CSIC-INTA), Carretera de Ajalvir, km. 4,
28850 Torrejón de Ardoz, Madrid, Spain. E-mail: Javiermt@cab.inta-csic.es

**Abstract:**

**Earth and Titan are two planetary bodies formed far from each other. Nevertheless the chemical composition of their atmospheres exhibits common indications of being produced by the accretion, plus ulterior *in-situ* processing of cometary materials. This is remarkable because while the Earth formed in the inner part of the disk, presumably from the accretion of rocky planetesimals depleted in oxygen and exhibiting a chemical similitude with enstatite chondrites, Titan formed within Saturn's sub-nebula from oxygen- and volatile-rich bodies, called cometesimals. From a cosmochemical and astrobiological perspective the study of the H, C, N, and O isotopes on Earth and Titan could be the key to decipher the processes occurred in the early stages of formation of both planetary bodies. The main goal of this paper is to quantify the presumable ways of chemical evolution of both planetary bodies, in particular the abundance of CO and $N_2$ in their early atmospheres. In order to do that the primeval atmospheres and evolution of Titan and Earth have been analyzed from a thermodynamic point of view. The most relevant chemical reactions involving these species and presumably important at their early stages are discussed. Then we have interpreted the results of this study in light of the results obtained by the Cassini-Huygens mission on these species and their isotopes. Given that H, C, N and O were preferentially depleted from inner disk materials that formed our planet, the observed similitude of their isotopic fractionation, and subsequent close evolution of Earth's and Titan's atmospheres points towards a cometary origin of Earth atmosphere. Consequently, our scenario also supports the key role of late veneers (comets and water-rich carbonaceous asteroids) enriching the volatile content of the Earth at the time of the Late Heavy Bombardment of terrestrial planets.**

## 1. Introduction

While the Earth formed in the inner part of the disk, presumably from the accretion of rocky planetesimals with a chemical similitude with enstatite chondrites (Wasson, 1988; 2000), Titan formed within Saturn's sub-nebula from volatile-rich bodies, called cometesimals (Alibert and Mousis, 2007). Despite their different origins, the chemical and isotopic composition of the atmospheres of Titan and Earth exhibit common features of being produced by the accretion, plus ulterior *in-situ* processing of cometary materials (see e.g. Owen and Niemann, 2009). This fact is quite remarkable because it could suggest that the volatiles were delivered from cometesimals formed in the outer solar system regions. A significant progress in our knowledge of Titan's atmosphere has been achieved from the Cassini-Huygens mission (Coustenis et al., 2009) that can be used to better understanding the evolution of volatile-rich atmospheres in planetary bodies. From a cosmochemical and astrobiological perspective the study of the isotopic fractionation of H, C, N, and O can be the key to decipher the processes occurred in the early stages of formation of both planetary bodies. These essential elements in organic chemistry were preferentially depleted from inner disk materials. Volatile-rich particles were overheated when falling towards the Sun. Consequently, volatile species were vaporized and recycled when the solar wind swept such gases to the outer disk. Phase transitions produced most of the isotopic exchanges because the heating of ices is needed to bring back the isotopes to the gaseous phase. 81P/Wild 2 materials have revealed that tiny minerals, plus ices and organics inherited these anomalies at the time of cometesimal formation (Brownlee et al., 2006). However, the recent discovery of ultra carbonaceous interplanetary dust particles (IDPs) suggests that the carbon and other abundances of biogenic elements in some outer bodies could have been far from the standard CI chondritic ratio (Duprat et al., 2010). Some of these volatile-rich objects escaped collisional processing, and subsequent aqueous alteration (Trigo-Rodríguez and Blum, 2009a).

The similitude between the volatile fraction of Earth and Titan points towards a major volatile enrichment of the Earth trough the called Late Heavy Bombardment (hereafter LHB) about 3.9 Ga ago. What can we consider as the more abundant chemical species building the primeval atmospheres of Earth and Titan? (For primeval we mean the post-accretionary atmosphere formed during the first hundreds million years, i.e., the Hadean for Earth's case). At that period we expect that several volatile species were obtained from thermal outgassing of chondritic material as a function of atmospheric temperature, pressure, and bulk composition. The chondrites can be considered the building blocks of the outer volatile fraction of Earth. Schaefer and Fegley (2007) obtained that the major outgassed volatiles from ordinary chondrite materials are $CH_4$, $H_2$, $H_2O$, $N_2$, and $NH_3$. These chemical species are the best candidates we have for the primeval constituents of the atmospheres of Earth as we know that ordinary and particularly enstatite groups of chondrites were major contributors to Earth's building blocks (Wasson, 2000). Schaefer and Fegley (2007) also demonstrated that CO is never the major C-bearing gas released during the metamorphism of chondritic meteorites as the released gases are leading to a more reducing atmosphere. Consequently CO and $CO_2$ probably are produced by secondary processes in a more oxidizing planetary environment, subject to the influx of larger amounts of organic matter, and volatiles as probably happened later on with the arrival of the late veneers to the Earth-Moon system 4 Ga ago (Gomes et al., 2005). Most probably the origin of $CO_2$ and CO in Titan was different because their initial forming blocks were rich in volatile compounds. Consequently, outgassing of the icy component of Titan's building blocks provided important amounts of $CO_2$, $H_2O$, $NH_3$, and $CH_4$ to its primeval atmosphere (Lunine et al., 2009). Despite that there is not a geological record of the primeval environment of Titan and Earth, in this paper we explore several reactions that were presumably important in both environments, but perhaps at different epochs. We will also address here two important issues:

a) The so-called CO deficiency issue in Titan: why CO is so underabundant in Titan while it is the second most abundant volatile in comets?

b) The role of molecular nitrogen ($N_2$) in the presumable escape of the Earth's and Titan's primeval atmospheres. Lichtenegger et al. (2010) has shown that N-rich atmospheres may be not stable due to the high Extreme Ultraviolet (EUV) flux of the young Sun.

## 2. Compositional comparison: main atmospheric species and isotope ratios

Current Earth and Titan's atmospheres are dominated by molecular nitrogen – Titan also contains $CH_4$ and $H_2$ as other secondary constituents –. The dominance of $N_2$ is not surprising as this molecule is extremely stable. $N_2$ has a triple bond (N≡N) with dissociation energy of 9.76 eV. This energy is higher than the C≡N bonds in HCN (9.58 eV) and CN (7.85 eV). The volume percent of $N_2$ in the atmosphere of Titan makes this world similar to Earth, and it seems to require a secondary processing of $N_2$ from $NH_3$ where photolysis, impacts and endogenic processes are some suggested candidates (Atreya et al., 2009). It is also remarkable that in the present Titan environment solar EUV photons and photoelectrons break the $N_2$ bond at high altitudes to produce HCN/nitriles. Similar processes could have occurred on Earth (more than 9 times closer to the Sun than Titan) during the Hadean eon of terrestrial evolution, particularly under a higher EUV flux than at present (Ribas et al., 2005; Lammer et al., 2008).

What clues can be obtained on the past of both objects from the major atmospheric species and their preserved isotope ratios in both planetary bodies? Besides the similarity of the $N_2$ dominated atmospheres in both planetary bodies, the isotopic ratios of C, H, and O are remarkably similar (see Fig. 1). This places Titan in the family of terrestrial planets. Just by comparing the measured isotopic ratios in Earth and Titan atmospheres with those measured in comets, new ideas on the origin of Earth's atmosphere and hydrosphere can be obtained. For example, it has been invoked that the D/H ratio is a central parameter to reconstruct the origin and pathways of water in the solar system (Greenberg, 1982; Eberhart et al., 1987; Delsemme, 2000). In Titan's atmosphere this ratio has been recently measured as $(1.32\pm0.15)\times10^{-4}$ (Bézard et al. 2007) that is smaller than the inferred $(2.09\pm0.45)\times10^{-4}$ from fitting acetylene bands (Coustenis et al., 2008). Interestingly the value by Bezard et al. is very close to the Standard Mean Ocean Water (SMOW) terrestrial value of $\sim1.6\times10^{-4}$ (see e.g. Meier & Owen, 1999). Despite of this similitude, the D/H ratio of Titan's building blocks could have been very different and enriched due to photochemical processes as previous works studying the D/H ratio in $CH_4$ in Titan have concluded (Pinto et al. 1986; Lunine et al. 1999; Cordier et al. 2008). This implies that the $CH_4$ in Titan's building blocks might not reflect the observed one and the fact that it is close to the SMOW value is just a coincidence. Another related point is that the D/H ratio in $H_2O$ in the building blocks of Enceladus has been measured to be cometary (Waite et al., 2009). Because formation scenarios predict that Titan's and Enceladus' building blocks share a common origin (Mousis et al. 2009), considering that the primordial D/H in water in Titan's building blocks should be similar and then enriched by a factor of 2 compared to SMOW sounds a good assumption.

On the other hand, the Titan ratio of $^{12}C/^{13}C=91.1 \pm 1.4$ is very close to the terrestrial inorganic Vienna-Pee Dee Belemnite standard value of 89.4 (Niemann et al., 2010). Moreover the $^{16}O/^{18}O$ ratio in Titan was recently estimated to be 346±110, while for Earth is 390 ± 144 (Nixon et al, 2009, Table 6), which is also close to the terrestrial value. An explanation for the slightly measured difference could be that the O ratios were fractionated in the nebula or during aqueous alteration of bodies formed nearby the snow line (Wasson, 2000), lately scattered due to the migration of the giant planets that produced the LHB. Finally, the Titan ratio of $^{14}N/^{15}N=56\pm8$ measured in HCN corresponds to an enrichment in $^{15}N$ of about five times the terrestrial ratio (Vinatier et al., 2007). In any case, we remark that the inferred ratio from $N_2$ is 167.7 ± 0.6 (Niemann et al., 2010) which is much closer to the terrestrial value of 272, just showing an enrichment of only 1.6 in the heavy isotope. The differences between the $^{14}N/^{15}N$ ratio inferred from $N_2$ and HCN can be explained by different photolysis rates between $^{14}N_2$ and $^{15}N^{14}N$ (see Liang et al., 2007).

## 3. Equilibrium thermodynamics calculation.

The H, C, N, and O isotope ratios of Titan, the terrestrial planets and comets have been measured by different techniques and compiled by different authors (see e.g. Atreya et al. 2009, for an updated review). We have compared these values in order to get clues on the origin of volatile compounds on Earth and Titan. A better understanding of the role of the presumable atmospheric constituents in the atmosphere of the Earth during the Hadean (see Table 1) would depend on a detailed study of their role in Titan.

In order to reveal the atmospheric evolution of the primeval atmospheres of Titan and Earth we have studied some chemical equilibrium reactions presumably important at the time of the primordial atmosphere. By using equilibrium thermodynamics we assume that the lower atmospheres of both bodies could have been in chemical equilibrium for most of their past. Our main goal has been to identify the most probable primordial composition of the Titan and Earth atmospheres, and quantify the presumable ways of chemical evolution.

Following this approach we analyze the evolution over long time scales of the main species. For that we use a simple model of independent linear equations for each species under study that takes into account chemical reactions of the form:

$$aA + bB \leftrightarrow cC + dD \qquad (1)$$

Every chemical reaction (1) has an equilibrium constant:

$$K = \frac{C^c \cdot D^d}{A^a \cdot B^b} \qquad (2)$$

that is related to the Gibbs free energy (G) as:

$$\Delta G = -RT \ln K$$

where R is the gas constant and T the temperature.

We note that by definition $\Delta G$ equals the work exchanged by the system with its surroundings during a reversible transformation from the same initial state to the final state. Therefore, we can study the evolution of different atmospheric compounds by considering the chemical reactions, the main species involved, and using the tabulated free energies of formation of these components. We have used the Gibbs energies of formation of the different compounds given in JANAF thermodynamic tables (1998). From them were built the ΔG and K values for the set of equations listed in table 2.

### 3.1. CO and $CO_2$

First we analyze the stability of CO and $CO_2$ in the primeval atmospheres using the reaction: $H_2$(g) + $CO_2$ (g) ↔ $H_2O$(g) + CO(g). From this reaction, taking into account the Gibbs free energies for the different components and assuming that: (i) the partial pressure of water was similar to the one observed on the present atmosphere of Earth ($pH_2O$~0.1 atm); and (ii) a partial pressure of $H_2$ within the range proposed in the literature (see Table 1), we obtain the ratio $pCO/pCO_2$ given in Figure 2. It seems evident that at temperatures lower than 600 K, the CO abundance is minor compared with that of $CO_2$. This should explain the low concentrations of CO in Titan. For Earth, it seems possible that the partial pressure of $H_2$ was higher in the past than nowadays because in presence of free Fe and a much lower $Fe^{3+}/Fe^{+2}$ ratio, the initial outgassed volatiles may have been in a much less oxidized form. Depending on a lower or higher abundance of $CO_2$ the primary iron-rich minerals could be respectively magnetite ($Fe_3O_4$) or siderite ($FeCO_3$) (Kasting, 2010). Both minerals have been identified in ancient soils called Banded Iron Formations (BIF) and recently analysed to constraint the $pCO_2$ atmospheric levels (Rosing et al., 2010). In any case, the hydrodynamic escape process is an important sink for $H_2$ that would be larger under high-temperature conditions (Tian et al., 2008). Under such a scenario perhaps the $pH_2$ reached its maximum value (about $10^{-2}$ atm) in the Hadean, decreasing with time until the present atmospheric value ($5\cdot10^{-8}$ atm). If CO was introduced from exogenous sources or produced in bolide plumes after giant impacts, as the temperature decreased the CO would be transformed progressively in $CO_2$.

### 3.2 $NH_3$ and $N_2$

Particularly, we analyzed the reaction: $3H_2(g) + N_2\ (g) \leftrightarrow 2\ NH_3(g)$ that provides clues on the evolution of the partial pressures on ammonia and molecular nitrogen in the post-accretionary atmosphere of Titan. It is remarkable that the production of ammonia under $pN_2{\sim}1$ requires significant abundance of $H_2$ (Fig. 3). For the range of temperatures assumed to be present in the early Titan, the $N_2$ originated by irradiation or impacts prevents an ulterior transformation to ammonia only for those models involving low partial pressures of $H_2$.

### 3.3 $CH_4$ and $CO_2$

Finally, to analyze the stability of $CH_4$ and $CO_2$ in both, Earth and Titan, we made equilibrium calculations for the reaction: $4H_2(g) + CO_2\ (g) \leftrightarrow 2\ H_2O(g) + CH_4(g)$ taking into account the Gibbs free energies for the different components and, again, assuming that the partial pressure of water was similar to that in the present atmosphere ($pH_2O{\sim}0.1$); and that the partial pressure of $H_2$ was inside the range usually accepted in the literature for the Earth (Table 1). The results are shown in Figure 4. From the derived $pCH_4/pCO_2$ ratio it is evident that the role of methane as greenhouse gas could have been important during the Hadean.

## 4. **Discussion: Titan versus primeval Earth**

Different instruments onboard the *Cassini-Huygens* mission obtained crucial measurements that have helped to our understanding of the isotopic ratios present in Titan's dense atmosphere. The new picture obtained from the data provides important constraints on the origin, and evolution of Titan, but also raises new questions about its amazing similarity with planet Earth. Some atmospheric evolutionary processes in Earth and Titan would have been similar, but other not, as consequence of very different radiation environments, equilibrium temperatures, endogenic processes, and impact rates. Differentiated isotope fractionation processes occurred in Titan's atmosphere would have been consequence of different mixing ratios, and solar flux compared to Earth. Despite of this, the similarity of the isotopic ratios of H, C, N, and O is remarkable. $H_2$ and $N_2$ species could have suffered a massive escape at early times of Titan's evolution. Note that Lichtenegger et al. (2010) have recently demonstrated that N-rich atmospheres may be not stable due to the high EUV flux of the young Sun. The Earth's atmosphere could have been lost before 4 Ga ago if was mainly composed by molecular nitrogen as nowadays. Similar problems could occur on Titan due to its low gravity field during the high EUV flux period of the young Sun (the first 200 - 500 Ga after the Sun arrived at the Zero Age Main Sequence (ZAMS)).

Based on the current data we propose here an scenario were the Earth partially lost its atmosphere 4 Ga ago and recovered it again during the LHB. A similar evolution of the volatiles in our planet and Titan can be suggested. The $^{15}N$ enrichment could be consequence of a significant amount of $N_2$ lost since the formation of the earliest atmosphere of Titan (Atreya et al., 2009; Lammer et al., 2008; Penz et al., 2005). It is also remarkable the close similitude for the D/H and $^{12}C/^{13}C$ ratios in both atmospheres (Fig. 1). The $^{16}O/^{18}O$ ratio is also close within the error suggesting that oxidized building blocks of Earth and Titan could have shared a common origin in the nebula. Consequently, most of the mentioned isotopic ratios suggest signatures inherited in Earth and Titan from volatiles preserved in comets, but other require additional modeling of the earliest atmospheric evolution of Titan and Earth. In particular nitrogen fixation by intense volcanic lightning in the early Earth (Navarro-González et al., 1998) could explain the specific nitrogen-dissimilarity in both planetary bodies.

Volatile-rich bodies were preferentially scattered from the outer solar system regions during the LHB period as a consequence of the giant planets migration invoked by the Nice model (Gomes et al., 2005). We expect a crucial role of these late veneers in the enrichment of the atmosphere of Earth, making it similar to that one of Titan. In this sense, the isotopic fractionations of H, C, N and O in Earth and Titan's atmospheres are similar to those expected for primordial volatile materials delivered from comets (Owen, 1982; Owen and Bar-Nun, 1995; Owen and Niemann, 2009). This is most probably a direct evidence of the important role of comets in the volatile enrichment of Earth previously claimed by several authors (see e.g. Oró, 1961; Oró et al., 1990; Owen and Bar-Nun, 1995; Delsemme, 2000). On the other hand, the different ways of trapping volatiles in planetesimals is a complicated issue that is out of the scope of this paper, but they have been widely discussed in Lunine et al. (2009).

Was water brought by comets or asteroids? This unsolved issue can have a no so complicated answer as in the outer part of the belt both definitions remain diffuse. Probably many bodies in that region accreted significant amounts of water ice and complex organics as suggests the recent discovery of Main Belt comets (Hsieh & Jewitt, 2006). In fact, it has been suggested that the precursor materials of carbonaceous chondrites possibly accreted higher amounts of volatiles than previously believed (Nuth, 2008; Trigo-Rodríguez & Blum, 2009a). If so, we cannot consider any group of carbonaceous chondrite, not even any combination of present meteorite types, as representative of the late veneers' population because such bodies have evolved with time towards higher compacted samples (Blum et al., 2006). We expect that the main population of impactors during the LHB was presumably formed by highly porous and easily disrupting volatile-rich bodies that particularly enriched the volatile-depleted content of terrestrial planets (Trigo-Rodríguez & Blum, 2009a). This is consistent with small asteroids suffering collisionally-induced compaction, brecciation and aqueous alteration of their forming minerals (Trigo-Rodríguez & Blum, 2009b).

In any case, we wish to remark that many issues remain open. As a recent evaluation of the noble gas signatures indicates, the necessary fraction of cometary matter brought during the LHB to account for the noble gas atmospheric inventory is $<1\%$ (Marty and Meibom, 2007). This amount is by far lower than the one predicted by Gomes et al. (2005) modeling ($\approx 50\%$). Ulterior simulations by O'Brien et al. (2006)

show that the mass fraction of material in the terrestrial planets from beyond 2.5 AU ranges from 1.6 to 38%, with a median value of 15%. As a fraction of planet mass, the median mass fraction of material delivered in the late veneer epoch was estimated to be of 0.7% (O'Brien et al., 2006). This value could be consistent with Marty and Meibom (2007) suggestions about the noble gas inventory, and could be enough to explain the water inventory. Despite of this we remark that the proportion of water delivered by comets or Kuiper Belt Objects (KBOs) could be much higher than the 10% assumed from carbonaceous chondrite meteorite analysis because they are the biased toughest samples arrived from largest and probably inhomogeneous bodies that 4 Gyrs ago could contain highest abundances of water before experiencing collisional processing (Trigo-Rodríguez & Blum, 2009a). An example is the evidence in carbonaceous groups of chondrites (like e.g. the CM water-rich group) of hydrated and unhydrated meteorites coming each one from different regions of a parent asteroid (Trigo-Rodríguez et al., 2006; Rubin et al., 2007). At the same time, the Earth's mantle is enriched in highly siderophiles elements relative to the expected from core formation (Morgan et al., 2001) which can be interpreted by adding about a 1% of late veneer material to the mantle long after core formation ceased (Drake and Righter, 2002; O'Brien et al., 2006).

The most abundant N-bearing molecules measured from remote observation of comets are $N_2$ and $NH_3$ with a maximum inferred abundance ratio relative to water of 1% and 1.5% respectively (Krueger & Kissel, 1987; Rauer, 2008). Other N-bearing molecules like $N_2$, HCN, $HC_3N$, $CH_3CN$, $NH_2CHO$, etc. could account by another additional ≈1% relative to water (Rauer, 2008). To this value we should add the N host in minerals, and the carbonaceous matrix that is particularly N-rich in CI and CM groups of chondrites (Rubin & Choi, 2009). Could this total 2-5% N content be enough to explain the present day 800 mbar of N in the terrestrial atmosphere?. We think it is possible, particularly taking also into account the continuous flux of volatile elements from IDPs, and meteoroids (see e.g. Anders, 1989), but to answer this point could be premature on the light of our current knowledge on the N abundance in comets. We remark how unknown is, for example, the amount of N host in cometary interiors. We hope that future Rosetta ESA mission experiments to measure the volatile elements contained in comet 67P/Churyumov-Gerasimenko could give additional clues to answer this question (e.g. Morse et al., 2009).

We see Titan as a natural oasis of remarkable astrobiological significance to understand the environment in which origin of life took place on Earth (Raulin et al. 2009). Figure 2 exemplifies that $H_2$-rich atmospheres tend to promote a higher abundance of CO relative to $CO_2$. Note that having in the early times a quick hydrodynamic escape will decrease the abundance of molecular hydrogen. This, together with the expected temperature decrease with time, indicates that the atmosphere should tend to have more $CO_2$ than CO (see Fig. 2). In any case, we obviously note that equilibrium calculations to decipher the persistence of CO or $CO_2$ species should be taken with caution. For example, Kasting (1989, 1997) proposed for the early Earth that the rapid weathering of the ejecta from frequent large impacts could have provided a sink for the atmospheric $CO_2$. In consequence, short-time scale stochastic events (eruptions, impacts, etc…) could produce abundances well out of chemical equilibrium. Note that the rapid conversion of CO into $CO_2$ and $CH_4$ predicted by our simple model relaxes the temperature constraints that were imposed to the satellite planetesimals in order to explain their current atmospheric compositions (Alibert and Mousis, 2007; Mousis et al., 2009), in particular the low abundance of CO in Titan's atmosphere.

On the other hand, our results suggest that ammonia and methane could have been present in the primeval atmospheres of Earth and Titan. Figures 3 and 4 exemplify that $H_2$-rich atmospheres tend to promote a higher abundance of $NH_3$ and $CH_4$. Both species probably played an important role as greenhouse gases. In fact, Sagan & Chyba (1997), attempting to solve the called faint young Sun paradox, proposed the formation of particles of organic polymers (tholins) by the action of UV light in an atmosphere with $CO_2/CH_4<1$. According to their model, methane and ammonia participated in a kind of feedback cycle, working together in two main ways: (i) methane's presence produced an organic haze due its capacity to polymerize and (ii) this haze can protect ammonia from UV photodissociation. In any case, the tendency of $NH_3$ and $CH_4$ to decompose in short time scales by solar irradiation has been the main criticism of the community against a strongly reducing Hadean terrestrial atmosphere. Current lifetime of atmospheric $CH_4$ in Titan is ~ 15 Ma so it must be continually resupplied from the interior. On Titan $CH_4$ complex photochemistry occurs with high efficiency as most hydrogen (H or $H_2$) escapes at the limiting diffusion velocity and negligible $CH_4$ is recycled by the reaction H + $CH_3$. The irreversible photolysis of $CH_4$ leads to the formation and deposition of $C_2H_x$ compounds on the surface. This behaviour is extremely interesting as Titan nowadays could have a similar chemistry to the one that the Earth had at the end of the Hadean eon. Suspended in the stratosphere, these polymers protect the lower atmosphere from UV photolysis acting as a screen capable to avoid the photodecomposition of these greenhouse gases. Initially the community was against the presence of other greenhouse gases as $NH_3$ or $CH_4$ based on the argument that these gases were easily photodissociated by UV light than $CO_2$.

Then, according to previous discussion, it seems that a plausible scenario to build life consists of a dense atmosphere where small particles like e.g. organic haze and meteoric metals (Sekine et al., 2003) could act as catalysts for the formation of more complex organic compounds from simple precursors such as $CO_x$ and $CH_4$, thus promoting increasing complexity (Trigo-Rodríguez et al., 2010). Obviously Titan's photochemistry (Coustenis et al., 2009) is nowadays exemplifying the complexity of this problem. We think that this enriching period was common to Earth and Titan, but with photochemical and geochemical differences inherent respectively to their different distances to the Sun, and surface minerals available after accretion.

## 5. Conclusions

We have analyzed the primeval atmospheres of Titan and Earth and their evolution from a thermodynamic point of view in order to quantify the presumable ways of chemical evolution of both planetary bodies, and, in particular to explain the abundances of CO and $N_2$ in their early atmospheres. The main conclusions of this work can be summarized as follows:

- Titan provides an extraordinary environment to better understand some of the chemical processes that lead to the apparition of life on Earth. Its atmosphere is a natural laboratory that, in many aspects, seems to have a strong similitude with our current picture of the pre-biotic atmosphere of Earth.
- For the range of $0.01<p(H_2)<1$ and from the derived $pCH_4/pCO_2$ ratio, the role of methane as greenhouse gas was probably important during the Hadean.

- The $CH_4$ computed abundances: $10^{-4} \leq pCH_4/pCO_2 \leq 1$ have UV shielding implications for stability and production of organics. Such balance can be additionally reinforced by destruction of $CO_2$.
- Current signatures in the atmospheres of Earth and Titan suggests that their early atmospheres were affected by a massive escape of $N_2$ under a strong EUV solar flux.
- The rapid conversion of CO into $CO_2$ and $CH_4$ predicted by our thermodynamic model relaxes the temperature constraints that are usually imposed to the satellite planetesimals in order to explain their current atmospheric compositions. This also should explain the deficit of CO in Titan.
- A $CH_4$- and HCN-rich scenario could act as catalysts for the formation of complex organic compounds during continuous ablation processes, from simple chemical precursors, thus promoting increasing complexity. Organic enrichment of the Hadean Earth surface as consequence of the LHB could be considered the first step of prebiotic evolution.
- Finally, given that light elements as e.g. H, C, N and O were preferentially depleted from inner disk materials that formed our planet, the isotopic similitude between these species in Earth and Titan is pointing towards a cometary origin of Earth atmosphere.

Among the future challenges to get new clues in this regard, the robotic exploration of Titan to study its surface and subsurface would answer some of the questions raised from the Cassini-Huygens scientific breakthrough.

## Acknowledgements

JMTR acknowledges financial support from Spanish National Plan of Scientific Research of MICINN under grant AYA2008-01839/ESP. He also recognizes the motivation received to study the early stages of Earth's atmosphere and the very suggestive teachings received from Prof. J. Oró and Prof. John T. Wasson (UCLA). This paper is dedicated to the memory of Prof. J. Oró. We also thank Dr. Olivier Mousis, an anonymous reviewer, and the volume editor Dr. Conor Nixon for their insightful suggestions.

# TABLES

Table 1. Partial pressure of presumable components of the Hadean terrestrial atmosphere.

| Gas | REFERENCES | | | | Comments |
|---|---|---|---|---|---|
| | Holland, 1984 | Kasting, 1993 | Sagan & Chyba 1997 | Sekine et al, 2003 | |
| $N_2$ | $0.1 \le pN_2 \le 1$ | $0.8 \le pN_2 \le 1$ | - | $pN_2 \approx 0.8$ (from Kasting) | Inactive and assumed to have similar abundance to the present atmosphere. |
| $CO_2$ | $10^{-5} \le pCO_2 \le 10^{-3}$ | $10^{-1} \le pCO_2 \le 10$ | probably $pCO_2 \le 10^{-1}$ | $pCO_2 \approx 10^{-1}$ (assumed from Kasting) | Farquhar & Wing (2003) propose an upper limit of $[CO_2] \le 0.5$ bar according to $\Delta^{33}S$ evidence. |
| CO | $10^{-8} \le pCO \le 10^{-6}$ | $10^{-1} \le pCO \le 1$ | - | $10^{-2} \le pCO \le 10^{-1}$ | minor relative to $CO_2$ |
| $H_2$ | $10^{-7} \le pH_2 \le 10^{-3}$ | $pH_2 \approx 3 \times 10^{-3}$ | - | $10^{-3} \le pH_2 \le 1$ | High differences depending on the source. |
| $NH_3$ | $pNH_3 \le 10^{-5}$ (if existed) | $10^{-6} \le pNH_3 \le 10^{-4}$ | $10^{-6} \le pNH_3 \le 10^{-4}$ | - | Proposed greenhouse to solve faint Sun dilemma. Its survival to UV light requires presence of $CH_4$. |
| $CH_4$ | $pCH_4 \le 10^{-2}$ | $10^{-5} \le pCH_4 \le 10^{-3}$ | $10^{-5} \le pCH_4 \le 10^{-4}$ | $pCH_4 \le 10^{-1}$ | Proposed greenhouse, whose polymerization produces UV protective hazes |
| $O_2$ | Negligible | - | - | - | Nearly absent as deduced from MIF of Sulfur isotopes |

Table 2. Compilation of the net ΔG, and K values for each reaction depending on temperature. Original thermodynamic data for every species were taken from JANAF (1998) obtained for a pressure of 1 bar. ***[Note for final edition that the columns should be aligned, and having the same size for each Temperature. Current display is a Word edition artefact]:***

| T (K) | 200 | 300 | 400 | 500 | 600 | 700 | 800 | 900 | 1000 | 1100 | 1200 | 1300 | 1400 | 1500 |
|---|---|---|---|---|---|---|---|---|---|---|---|---|---|---|
| $H_2(g) + CO_2(g) \leftrightarrow H_2O(g) + CO(g)$ | | | | | | | | | | | | | | |
| ΔG | 32793 | 28566 | 24436 | 20474 | 16689 | 13068 | 9593 | 6249 | 3021 | -107 | -3146 | -6106 | -8998 | -11828 |
| K | $2.72\times10^{-9}$ | $1.06\times10^{-5}$ | $6.4\times10^{-4}$ | $7.27\times10^{-3}$ | $3.52\times10^{-2}$ | 0.106 | 0.236 | 0.434 | 0.695 | 1.012 | 1.371 | 1.759 | 2.166 | 2.581 |
| $3H_2(g) + N_2(g) \leftrightarrow 2\,NH_3(g)$ | | | | | | | | | | | | | | |
| ΔG | -51358 | -32366 | -11882 | 9600 | 31758 | 54380 | 77324 | 100494 | 123820 | 147250 | 170746 | 194282 | 217836 | 241392 |
| K | $2.58\times10^{13}$ | $4.31\times10^{5}$ | 35.6 | 0.099 | $1.72\times10^{-3}$ | $8.75\times10^{-5}$ | $8.94\times10^{-6}$ | $1.47\times10^{-6}$ | $3.41\times10^{-7}$ | $1.02\times10^{-7}$ | $3.7\times10^{-8}$ | $1.6\times10^{-8}$ | $7.5\times10^{-9}$ | $3.9\times10^{-9}$ |
| $4H_2(g) + CO_2(g) \leftrightarrow 2\,H_2O(g) + CH_4(g)$ | | | | | | | | | | | | | | |
| ΔG | -129608 | -113224 | -95181 | -75904 | -55719 | -34869 | -13521 | 8198 | 30198 | 52407 | 74772 | 97255 | 119823 | 142454 |
| K | $3.67\times10^{22}$ | $6.09\times10^{14}$ | $8.8\times10^{9}$ | $4.05\times10^{6}$ | $1.44\times10^{4}$ | $1.89\times10^{2}$ | 6.09 | 0.37 | $3.7\times10^{-2}$ | $5.23\times10^{-3}$ | $9.91\times10^{-4}$ | $2.35\times10^{-4}$ | $6.7\times10^{-5}$ | $2.24\times10^{-5}$ |

# FIGURES

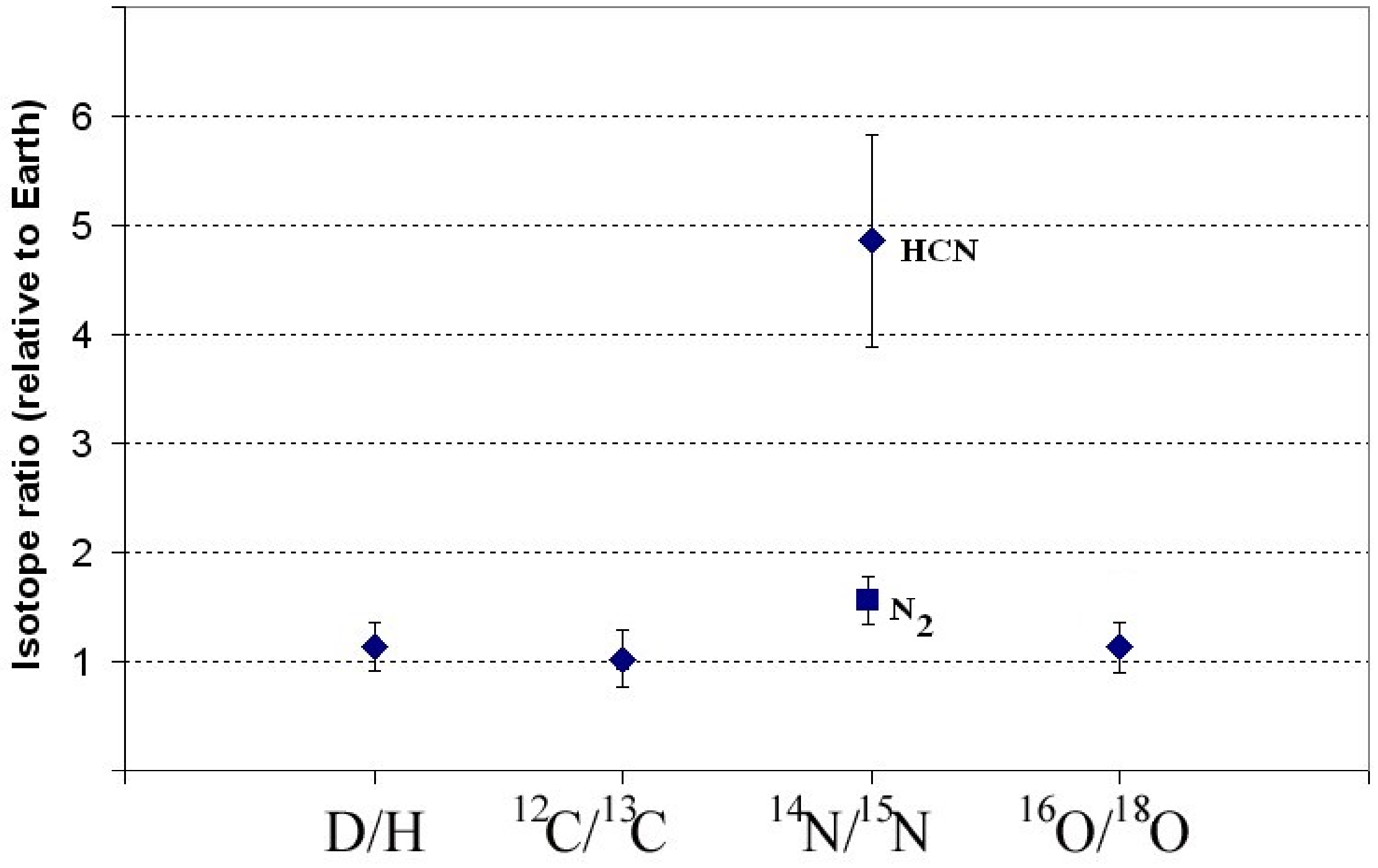

Figure 1. Titan isotope ratios of the light elements relative to Earth (value=1). Note that for $^{14}N/^{15}N$ are given the values inferred from HCN and $N_2$. For references, and additional discussion please see the text.

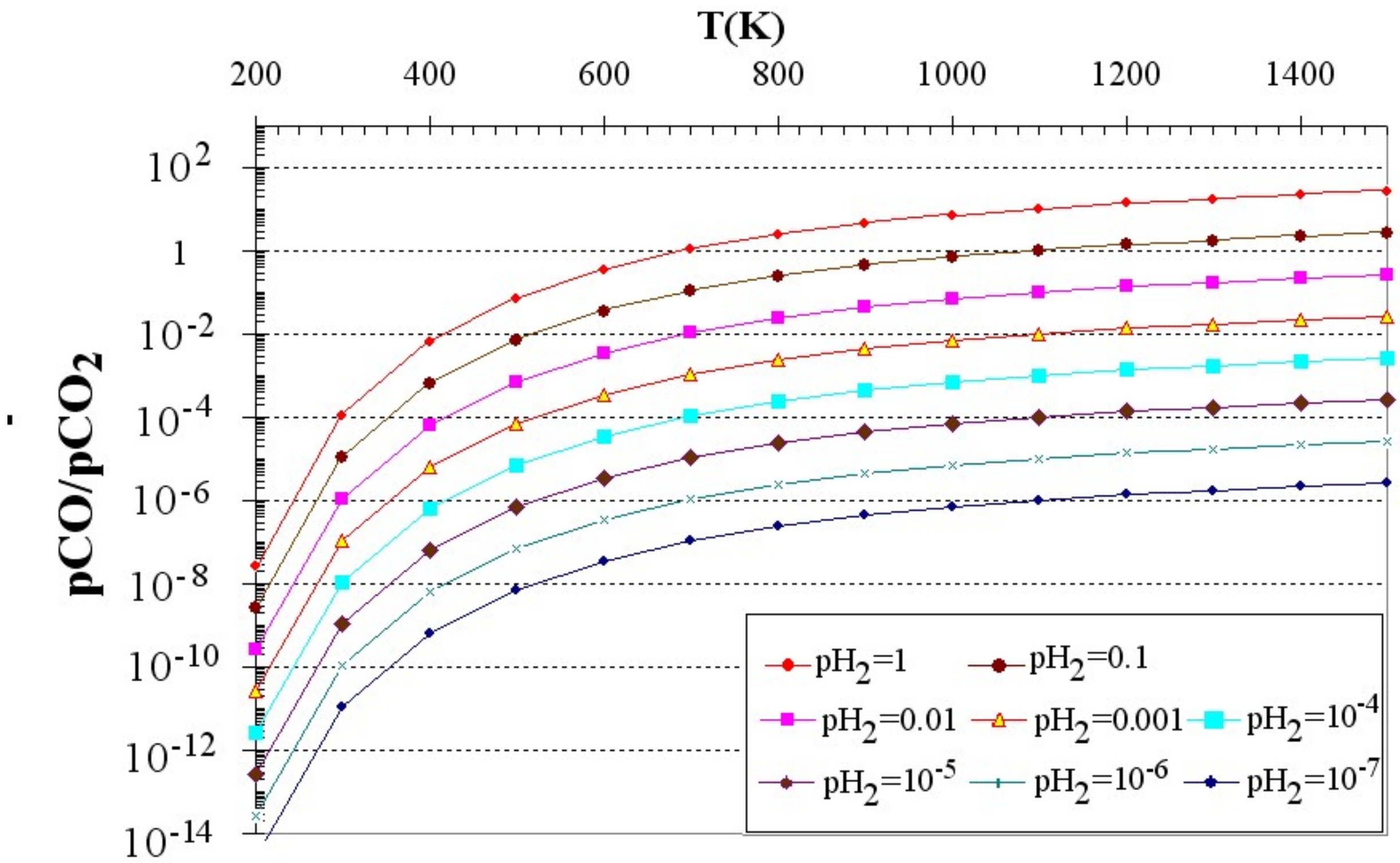

Figure 2. The ratio of partial pressures of CO and $CO_2$ plotted for post-accretionary conditions in Titan. Every track shows the evolution in the $pCO/pCO_2$ ratio as a function of temperature and the partial pressure of $H_2$ given in the box.

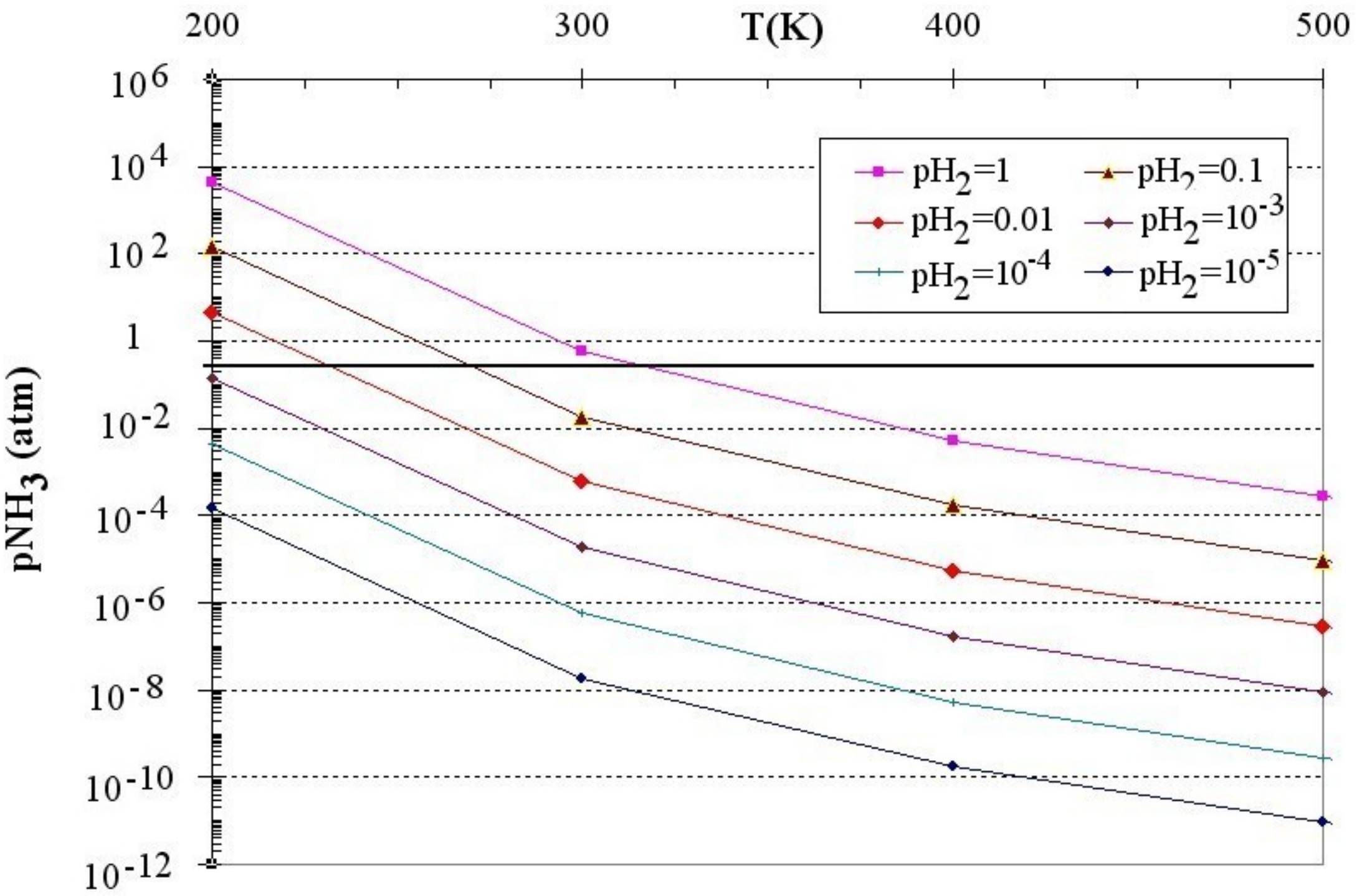

Figure 3. The partial pressure of $NH_3$ ($pNH_3$) plotted for post-accretion conditions. Every track shows the evolution in the partial pressure of ammonia as a function of temperature and the partial pressure of $H_2$ ($pH_2$) given in the box.

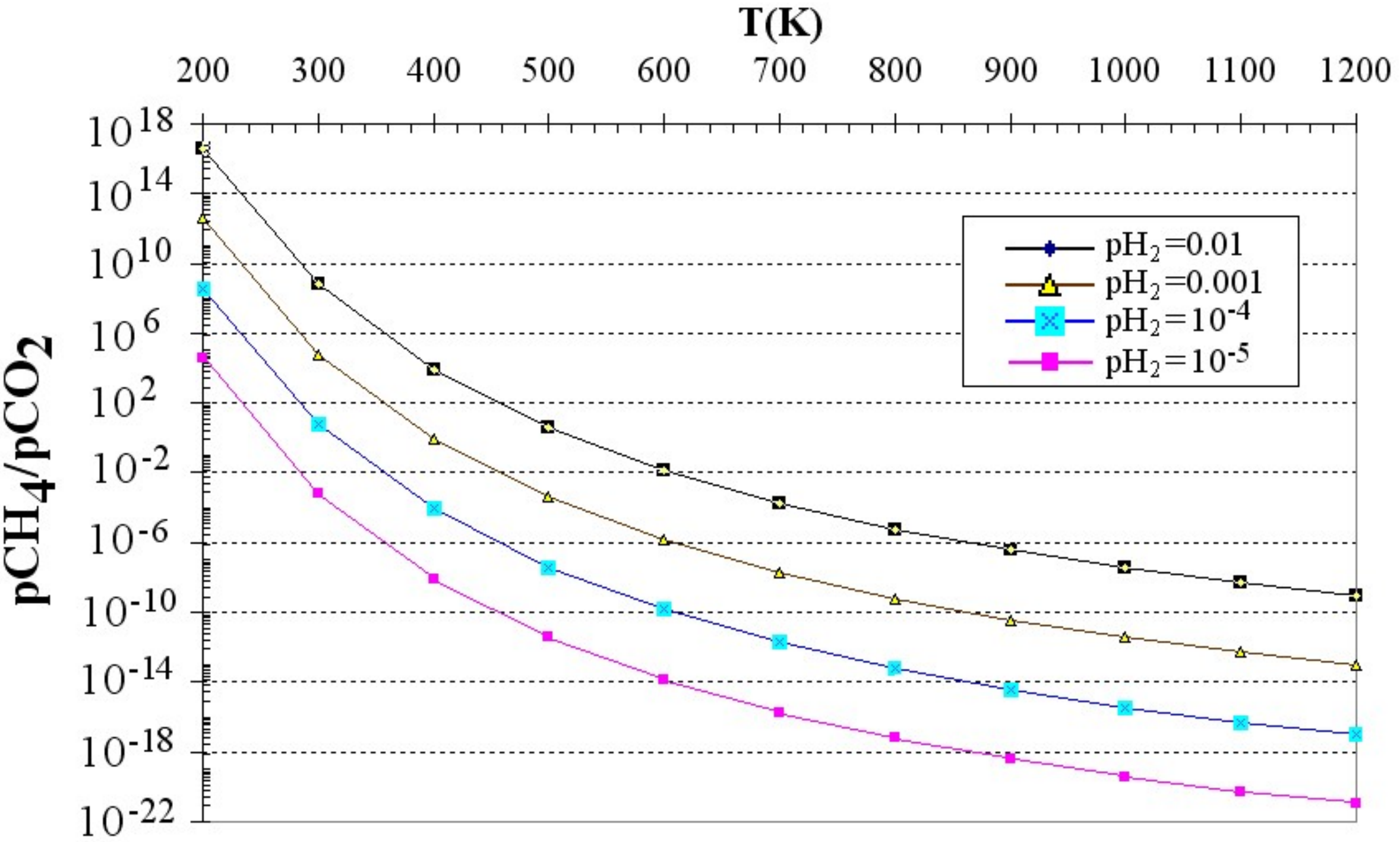

Figure 4. The partial pressure of $CH_4$ relative to that of $CO_2$ plotted for post-accretionary conditions. Every track shows the evolution in the $pCH_4/pCO_2$ ratio as a function of temperature and the partial pressure of $H_2$ given in the box.